\documentclass[a4paper,11pt]{article}
\usepackage{mathrsfs}
\usepackage{authblk}
\usepackage{tipa}
\usepackage{graphics}
\usepackage{amsmath}
\usepackage{amssymb}
\usepackage{bm}
\usepackage{amsfonts}
\usepackage{ntheorem}
\usepackage[colorlinks, linkcolor=red, anchorcolor=blue, citecolor=blue]{hyperref}
\usepackage{feynmf}
\usepackage{extarrows}
\usepackage{slashed}
\usepackage{graphicx}
\usepackage{appendix}
\usepackage{indentfirst}
\usepackage{color}
\author[1,2]{Debajyoti Sarkar\thanks{debajyoti.sarkar@physik.uni-muenchen.de}}
\author[3]{Xiao Xiao\thanks{xx2146@columbia.edu}}
\affil[1]{\emph{Department of Physics and Astronomy, Lehman College of the CUNY, Bronx NY 10468, USA}}
\affil[2]{\emph{Arnold Sommerfeld Center, Ludwig-Maximilians-University, Theresienstr. 37, 80333 M\"{u}nchen, Germany}}
\affil[3]{\emph{Physics Department and Institute for Strings, Cosmology and Astroparticle Physics}\\
\emph{Columbia University, New York, NY 10027, USA}}

\begin{document}

\date{}

%%%%%%%%%%%%%%%%%%%%%%%%%%%%%%%%%%%%%%%%%%%%%%%%%%%%%%%%%%%%%%
\title{\textbf{Holographic Representation of Higher Spin Gauge Fields}}
%%%%%%%%%%%%%%%%%%%%%%%%%%%%%%%%%%%%%%%%%%%%%%%%%%%%%%%%%%%%%%

\maketitle
\vspace{-10mm}
\begin{flushleft}
LMU-ASC 66/14
\end{flushleft}

\abstract{Extending the results of \cite{Heem}, \cite{KLRS} on the holographic representation of local gauge field operators in anti de Sitter space, here we construct the bulk operators for higher spin gauge fields at the leading order in $\frac{1}{N}$ expansion. Working in holographic gauge for higher spin gauge fields, we show that gauge field operators with integer spin $s>1$ can be represented by an integration over a ball region, which is the interior region of the spacelike bulk lightcone on the boundary. The construction is shown to be AdS-covariant up to gauge transformations, and the two-point function between higher spin gauge fields and boundary higher spin current exhibit singularities on both bulk and boundary lightcones. We also comment on possible extension to the level of three-point functions and carry out a causal construction for higher spin fields in de Sitter spacetime.

%%%%%%%%%%%%%%%%%%
\section{Introduction}
%%%%%%%%%%%%%%%%%%

It is well-known through the AdS/CFT correspondence \cite{Malda} that a strongly-coupled conformal field theory (CFT) with a large number of colors in $d$ dimensions is dual to a semiclassical gravity theory in $d+1$ dimensional anti de Sitter spacetime (AdS), which is a local field theory. In \cite{Wit} and \cite{GKP}, the so called GKPW prescription, explicit operator dictionaries were constructed that relates near-boundary bulk fields to operators in boundary CFT. To probe deeper, one has to identify the local operators deep in AdS with non-local operators in the boundary CFT. In this prescription, the non-normalizable part of the bulk field acts as the source to the boundary operators and using standard bulk to boundary propagators, the on-shell bulk fields are constructed. On the other hand, one can also take a slightly different route which involves identifying the normalizable part of the bulk field with the expectation value of the boundary operator itself (this can be done in Lorentzian-signature AdS/CFT as opposed to the Euclidean case). It is then a natural question to ask whether an identification exists at the (boundary) operator- (bulk) field level. For scalar field at leading order in large-$N$ expansion this was done in \cite{BDH}-\cite{Bena}, and was further refined in \cite{HKLL}. The construction was carried out in \cite{KLL} including bulk interaction at order $\frac{1}{N}$, which also explained the equivalence between two different approaches: one where we start from the CFT side and one where the local bulk operators are constructed using bulk equations of motion \cite{Heemskerk:2012mn}. In what follows, we implement the above-mentioned `normalizable mode' prescription to study these local bulk operators from the CFT side.

In \cite{Heem} and \cite{KLRS}, local operators with integer spins, especially gauge fields with zero mass such as bulk photon field and graviton field were constructed. The construction was shown to be AdS covariant, with two point functions between bulk gauge fields and boundary currents having bulk and boundary light-cone singularities. Causality was shown to be respected by gauge-invariant operators such as electro-magnetic field strength and Weyl tensor.\footnote{Here by (micro-) causality we will simply mean that the corresponding operator commutes with any other (bulk or boundary) local operator at spacelike separation.}

In this paper, we extend the construction to higher spin gauge fields with $s>2$. Even though interacting higher spin gauge theory suffers from various no-go relations in flat space \cite{Weinberg:1964ew}-\cite{Porrati:2008rm}, in AdS they can be well defined \cite{Vasiliev:1990en}, \cite{Vasiliev:2003ev}. Higher spin gauge field has recently attracted attention since a certain consistent higher spin gauge theory with interactions in AdS space \cite{Va} is conjectured \cite{KP} to be dual to free SO(N) vector model. Furthermore, the analytic continuation of this duality was proposed as a realization of the idea of dS/CFT in \cite{dSCFT}, \cite{AHS}. Recently \cite{Xiao:2014uea} discussed the smearing construction of scalars and spin- $s$ gauge fields in dS.\footnote{Here and afterwards, we will simply assume some form of dS/ CFT correspondence in its flat slicing (which can be generalized to the global coordinates). The bulk correlators are to be evaluated in Hartle-Hawking vacuum which when taken to the boundary, correspond to the Euclidean CFT correlation functions.} However the spin- $s$ gauge field construction in \cite{Xiao:2014uea} uses the normalizable boundary condition of AdS and analytically continues the result to the dS space. Such construction is manifestly acausal and kills either the positive or negative frequency modes of the bulk field. In the present paper, we improve the situation by including both the modes, which for AdS corresponded to both the normalizable and non-normalizable modes. This could be done by first considering the AdS bulk operators in terms of their boundary values on a cut-off surface in the bulk \cite{Sarkar:2014jia} and then analytically continuing the result to the dS space and subsequently taking the boundary limits. 

Another motivation to study bulk locality in the presence of bulk higher spin fields is intimately tied to the understanding of the required CFT properties for it to have a local bulk dual. The smearing construction from a point of view of microcausality is quite tailor-made to address such questions. A recent take towards answering such questions comes from the work of \cite{Heemskerk:2009pn}-\cite{Bekaert:2014cea}. The basic conjecture is that for a CFT to have a local bulk dual, all the single trace operators of spin greater than two should have parametrically large dimensions. This is a statement to be made at the level of four-point functions where we have the notion of locality of interactions and where interactions with spin- $s$ gauge fields are relevant in the bulk. It will be interesting to therefore see how does this constraint show up from the perspectives of microcausality. Here we don't study the four-point correlators and leave it for future studies. A starting calculation to fully appreciate the complexity of the problem, seems to simply work with a bulk scalar at this $\mathcal{O}(1/N^2)$ order \cite{toappear}.

To construct gauge fields with higher spin, we work in holographic gauge (to be discussed shortly), in which the calculation is simplified. We show that as in the cases for massless spin-1 and spin-2 fields, one can construct local spin- $s$ field in AdS bulk as a non-local operator smeared over certain region on the boundary; for $s>1$ the smearing function has support inside the bulk lightcone, while for $s=1$ the support is on the intersection between the bulk lightcone and the boundary (i.e. over a boundary ball and a spherical region respectively). Applying AdS isometries on the bulk operators generically brings the field out of the holographic gauge, but it is shown that one can always do a gauge transformation to bring the field back to holographic gauge, thus establishing the AdS covariance of the construction. Two-point functions of higher-spin fields and currents are calculated and shown to possess the singularity structure compatible with microcausality.

The plan of the paper is the following: In section \ref{sec:2} we briefly review the holographic construction of bulk scalar field, vector gauge field and graviton field, at the level of $N^0$. In section \ref{sec:3} we extend the construction to spin- $s$ gauge field and get an explicit smearing function. In section \ref{sec:4} we check the AdS covariance of the construction. It is shown that the construction, which is done in holographic gauge, is not covariant under AdS isometries, but is covariant up to a gauge transformation. In section \ref{2pfn} we look at two-point functions between a bulk gauge field and a boundary current where it is shown that besides the singularity on the bulk lightcone, singularity generically arises also on boundary lightcone, thus signaling non-locality for gauge fields. In section \ref{sec:6} we study the three-point function of higher spin fields using the leading order smearing function and show that similar bulk and boundary lightcone singularities arise there as well. Section \ref{hs_ds} is devoted to the higher spin field construction in de Sitter space. We conclude in section \ref{sec:conclusion} with some comments on bulk locality in the contexts of higher spins and provide prospective future directions. Previous studies of writing AdS$_4$ higher spin gauge fields in terms of boundary fields at all order in $1/N$ expansions (albeit, they were collective fields made out of $O(N)$ valued boundary fields) appeared in \cite{Koch:2010cy} and \cite{Koch:2014aqa}. The present construction seems much simpler and manifestly holographic.

%%%%%%%%%%%%%%%%%%%%%%%%%%%%%%%%%%%%%%%%%%%%%%%%%%%%%%
\section{Holographic representation of scalar, vector and tensor fields}\label{sec:2}
%%%%%%%%%%%%%%%%%%%%%%%%%%%%%%%%%%%%%%%%%%%%%%%%%%%%%%

In this section we briefly review the construction of local spin-0 and massless spin-1, spin-2 fields in AdS$_{d+1}$/CFT$_d$. We work in Poincar\'{e} patch with metric (taking AdS radius $R_{AdS}$ to be 1)
\begin{equation}\label{AdS_PP}
  ds^2=\frac{1}{z^2}\left( -dt^2+dz^2+d\textbf{x}^2 \right)
\end{equation}
A bulk scalar with dimension $\Delta=\frac{d}{2}+\sqrt{m^2+\left( \frac{d}{2} \right)^2}\equiv \nu+\frac{d}{2}$ can be constructed via summing over all normalizable modes in the bulk \cite{HKLL}:
\begin{equation}
  \Phi\left( t,z,\textbf{x} \right)=\int_{|\omega|>|\textbf{k}|}d\omega d^{d-1}\textbf{k} a_{\omega\textbf{k}}e^{-i\omega t}e^{i\textbf{k}\cdot\textbf{x}}z^{\frac{d}{2}}J_\nu\left( z\sqrt{\omega^2-\textbf{k}^2} \right)
  \label{}
\end{equation}
with the mode related to a boundary local operator by 
\begin{equation}
  a_{\omega\textbf{k}}=\frac{2^\nu\Gamma(\nu+1)}{(2\pi)^d\left( \omega^2-\textbf{k}^2 \right)^{\frac{\nu}{2}}}\int dt d^{d-1}\textbf{x} e^{i\omega t}e^{-i\textbf{k}\cdot\textbf{x}}\mathcal{O}(x)
  \label{}
\end{equation}
Putting $a_{\omega\textbf{k}}$ into the mode sum one obtains the representation of a local scalar field in AdS bulk as an integral on the boundary with compact support, via complexifying the boundary spatial coordinates.
\begin{equation}
  \Phi(t,z,\textbf{x})=\frac{\Gamma(\Delta-\frac{d}{2}+1)}{\pi^{\frac{d}{2}}\Gamma(\Delta-d+1)}\int_{t'^2+\textbf{y}'^2<z^2}dt' d^{d-1}\textbf{y}'\left( \frac{z^2-t'^2-\textbf{y}'^2}{z} \right)^{\Delta-d}\mathcal{O}(t+t',\textbf{x}+i\textbf{y}')
  \label{}
\end{equation}
For $\Delta>d-1$ the integral is well-defined, but it diverges for $\Delta=d-1$ which is a tachyon with mass $m^2=1-d$. The construction in this case was carried out in \cite{KLRS}, where it turns out that the integration domain is on the sphere $S^{d-1}$ on which the bulk lightcone intersects with the boundary:
\begin{equation}
  \Phi(t,z,\textbf{x})=\frac{1}{Vol(S^{d-1})}\int_{t'^2+\textbf{y}'^2=z^2}dt'd^{d-1}\textbf{y}'\mathcal{O}(t+t',\textbf{x}+i\textbf{y}')
  \label{}
\end{equation}
where $Vol(S^{d-1})=\frac{2\pi^{\frac{d}{2}}}{\Gamma(\frac{d}{2})}$ is the surface area of a $(d-1)$ sphere. 

The construction for scalars are very useful for the construction of gauge fields; the case for gauge fields (in holographic gauge) can always be reduced to the scalars with certain mass and be directly constructed with the help of the results discussed above for scalars.

Let's start from massless spin-one field. In this case, the source-free bulk Maxwell equation
\begin{equation}
  \nabla_M F^M_{~ ~N}=0
  \label{}
\end{equation}
can be solved in holographic gauge
\begin{equation}
  A_z=0
  \label{}
\end{equation}
One can further impose that
\begin{equation}
  \partial_\mu A^\mu=0
  \label{}
\end{equation}
which can be thought of as the conservation of the boundary current $\partial_\mu A^\mu\sim \partial_\mu j^\mu=0$. Here the Greek index $\mu$ runs from 0 to $d-1$ of the boundary coordinates, while the capital letters $M$ and $N$ run from 0 to $d$, including the radial or holographic coordinate $z$. The bulk equation for $A_\mu$ becomes
\begin{equation}
  \partial_\nu\partial^\nu A_\mu+z^{d-3}\partial_z\left( z^{3-d}\partial_z A_\mu \right)=0
  \label{}
\end{equation}
Defining 
\begin{equation}
  \Phi_\mu\equiv z A_\mu
  \label{}
\end{equation}
the equation for $A_\mu$ can be written as the free-wave equation for a multiplet of scalars with mass squared $m^2=1-d$
\begin{equation}
  \partial_\alpha\partial^\alpha\Phi_\mu+z^{d-1}\partial_z\left( z^{1-d}\partial_z\Phi_\mu \right)+\frac{d-1}{z^2}\Phi_\mu=0
  \label{}
\end{equation}
Thus the gauge field $A_\mu$ can be constructed via the construction of scalars $\Phi_\mu$:
\begin{equation}
  A_\mu(t,z,\textbf{x})=\frac{1}{z}\Phi_\mu(t,z,\textbf{x})=
  \frac{1}{Vol(S^{d-1})}\frac{1}{z}\int_{t'^2+\textbf{y}'^2=z^2}dt'd^{d-1}\textbf{y}'j_\mu(t+t',\textbf{x}+i\textbf{y}')
  \label{}
\end{equation}

The construction for graviton is similar. Working in holographic gauge
\begin{equation}
  h_{zz}=h_{z\mu}=0
  \label{}
\end{equation}
one solves the bulk equation for linearized graviton propagation in AdS space
\begin{equation}
  \nabla_Q\nabla^Q h_{MN}-2\nabla_Q\nabla_M h^Q_{~N}+\nabla_M\nabla_N h^Q_{~Q}-2d h_{MN}=0
  \label{}
\end{equation}
One can further impose the conditions
\begin{equation}
  h^\alpha_{~\alpha}=0~,~ \partial_\mu h^{\mu\nu}=0
  \label{}
\end{equation}
because of the conservation of the boundary currents and tracelessness of the stress tensor
\begin{eqnarray}
  & \partial_\mu T^{\mu\nu}=0\\
  & \partial_\mu\left( x_\nu T^{\mu\nu} \right)=T^\nu_{~\nu}=0
  \label{}
\end{eqnarray}
The $\mu\nu$ component of the bulk equation then gives
\begin{equation}
  \partial_\alpha\partial^\alpha h_{\mu\nu}+\partial_z^2 h_{\mu\nu}+\frac{5-d}{z}\partial_z h_{\mu\nu}-\frac{2(d-2)}{z^2}h_{\mu\nu}=0
  \label{}
\end{equation}
Defining a multiplet of scalars
\begin{equation}
  \Phi_{\mu\nu}\equiv z^2 h_{\mu\nu}
  \label{}
\end{equation}
we have the equation for $\Phi_{\mu\nu}$ as an equation for massless scalars\footnote{Similar relations between massless free scalars and the transverse traceless modes of the gravitons (as well as of its super-partners gravitino and gravi-photons, if present) were recognized very early in e.g. \cite{Bianchi:2000sm}.}
\begin{equation}
  \partial_\alpha\partial^\alpha\Phi_{\mu\nu}+z^{d-1}\partial_z\left( z^{1-d}\partial_z\Phi_{\mu\nu} \right)=0
  \label{}
\end{equation}
Therefore the bulk graviton can be represented as
\begin{equation}
  h_{\mu\nu}(t,z,\textbf{x})=
  \frac{\Gamma(\Delta-\frac{d}{2}+1)}{\pi^{\frac{d}{2}}\Gamma(\Delta-d+1)}\frac{1}{z^2}\int_{t'^2+\textbf{y}'^2<z^2}dt' d^{d-1}\textbf{y}'\left( \frac{z^2-t'^2-\textbf{y}'^2}{z} \right)^{\Delta-d}T_{\mu\nu}(t+t',\textbf{x}+i\textbf{y}')
  \label{}
\end{equation}

%%%%%%%%%%%%%%%%%%%%%%%%%%%%%%%%%%%%%%%%%%%%%%%
\section{Holographic representation of massless spin- $s$ field}\label{sec:3}
%%%%%%%%%%%%%%%%%%%%%%%%%%%%%%%%%%%%%%%%%%%%%%%

In this section we carry out the construction for a general integer spin gauge field in $AdS_{d+1}$ in terms of smeared local operators in the field theory.

As is well known, massless gauge fields in AdS are represented by totally symmetric rank-$s$ tensor $\Phi_{M_1\dots M_s}$ satisfying double-tracelessness conditions.
\begin{align}
   \Phi^{MN}_{~ ~ ~ ~MN M_5\dots M_s}=0
  \label{}
\end{align}
The linearized equation of spin- $s$ gauge field on AdS$_{d+1}$ is \cite{Va}, \cite{Mikh}, \cite{GY}, 
%fixed
\begin{equation}
  \nabla_N\nabla^N\Phi_{M_1\dots M_s}-s\nabla_N\nabla_{M_1}\Phi^N_{~M_2\dots M_s}+\frac{1}{2}s(s-1)\nabla_{M_1}\nabla_{M_2}\Phi^N_{~N\dots M_s}-2(s-1)(s+d-2)\Phi_{M_1\dots M_s}=0
  \label{}
\end{equation}
This equation is invariant under the gauge transformation
\begin{equation}
  \Phi_{M_1\dots M_s}\rightarrow \Phi_{M_1\dots M_s}+\nabla_{M_1}\Lambda_{M_2\dots M_s}~,~ \Lambda^N_{~ ~NM_3\dots M_s}=0
  \label{}
\end{equation}
We would like to work in holographic gauge in which all the $z$-components of the gauge field vanishes, so we shall generalize the holographic gauge from vector and rank-2 tensor to rank-$s$ tensor
\begin{equation}
  \Phi_{z\dots z}=\Phi_{\mu_1 z\dots z}=\dots =\Phi_{\mu_1\dots\mu_{s-1}z}=0
  \label{}
\end{equation}
One can always make this gauge choice because given a general field $\Phi^{(0)}_{M_1\dots M_s}$, one can perform a gauge transformation $\Phi\rightarrow \Phi+\nabla\Lambda$ so that
%fixed last two
\begin{align}
  & \Phi^{(0)}_{z\dots z}+\nabla_z\Lambda_{z\dots z}=0\\  
  & \Phi^{(0)}_{\mu_1 z\dots z}+\nabla_{z}\Lambda_{\mu_1z\dots z}=0\\
  & \dots\nonumber\\
  & \Phi^{(0)}_{\mu_1\dots \mu_{s-1} z}+\nabla_z\Lambda_{\mu_1\dots \mu_{s-1}}=0
  \label{}
\end{align}
The number of gauge parameters are just right to satisfy the set of equations and fix the holographic gauge for generic spin- $s$ gauge field.

The bulk gauge field is dual to totally symmetric, traceless, conserved rank-$s$ tensor on the boundary:
%fixed
\begin{equation}
  \mathcal{O}^\nu_{~\nu \mu_3\dots\mu_s}=0~,~\partial_\nu \mathcal{O}^\nu_{~\mu_2\dots\mu_s}=0
  \label{}
\end{equation}
In holographic gauge, only three components of the bulk equation are non-trivial, they are
\newline
for $zz\mu_3\dots\mu_s$ components:
\begin{equation}
  \frac{s(s-1)}{2}\partial_z^2\Phi^\alpha_{~\alpha\mu_3\dots\mu_s}+\left( \frac{s(s-1)(2s-3)}{2}-s \right)\frac{1}{z}\partial_z\Phi^\alpha_{~\alpha\mu_3\dots\mu_s}+\left( 2-s(s-1)+\frac{s!}{2(s-4)!} \right)\frac{1}{z^2}\Phi^\alpha_{~\alpha\mu_3\dots\mu_s}=0
  \label{}
\end{equation}
%fixed
for $z\mu_2\dots\mu_s$ components:
\begin{equation}
  \left( 2-s(s-1) \right)\frac{1}{z}\partial_\alpha\Phi^\alpha_{~\mu_2\dots\mu_s}-s\partial_z\partial_\alpha\Phi^\alpha_{~\mu_2\dots\mu_s}+\frac{s(s-1)}{2}\left( \partial_z\partial_{\mu_1}\Phi^\alpha_{~\alpha\mu_2\dots\mu_s}+\frac{s-1}{z}\partial_{\mu_1}\Phi^\alpha_{~\alpha\mu_2\dots\mu_s} \right)=0
  \label{}
\end{equation}
%fixed
for $\mu_1\dots\mu_s$ components:
\begin{align}
  &  z^2\left( \partial_z^2+\partial_\alpha\partial^\alpha \right)\Phi_{\mu_1\dots\mu_s}+(2s+1-d)z\partial_z\Phi_{\mu_1\dots\mu_s}+2(s-1)(2-d)\Phi_{\mu_1\dots\mu_s}-s\partial_{\mu_1}\partial_\alpha\Phi^\alpha_{~\mu_2\dots\mu_s}\nonumber\\
  &+\frac{1}{2}s(s-1)\left( \partial_{\mu_1}\partial_{\mu_2}\Phi^\alpha_{~\alpha\mu_3\dots\mu_s}-g_{\mu_1\mu_2}z\partial_z\Phi^\alpha_{~\alpha\mu_3\dots\mu_s}-(s-2)g_{\mu_1\mu_2}\Phi^\alpha_{~\alpha\mu_3\dots\mu_s}-\sum^s_{i=3}g_{\mu_1\mu_i}\Phi^\alpha_{~\alpha\mu_2\dots\mu_{i-1}\mu_{i+1}\dots\mu_s} \right)\nonumber\\
  &=0
  \label{}
\end{align}
and all the other components with more than two $z$'s vanish trivially.

Since the boundary currents are conserved and traceless, we can consistently set
\begin{align}
  & \Phi^\alpha_{~\alpha\mu_3\dots\mu_s}=0\\
  & \partial_\alpha\Phi^\alpha_{~\mu_2\dots\mu_s}=0
\end{align}
and we have the $\mu_1\dots\mu_s$ component of the equation as
\begin{equation}
  \left( \partial_z^2+\partial_\alpha\partial^\alpha \right)\Phi_{\mu_1\dots\mu_s}+\frac{2s+1-d}{z}\partial_z\Phi_{\mu_1\dots\mu_s}+\frac{2(s-1)(2-d)}{z^2}\Phi_{\mu_1\dots\mu_s}=0
  \label{}
\end{equation}
To turn the problem of constructing spin- $s$ gauge field into constructing sclars, we define\footnote{Just like in spin-1 and spin-2 cases, this amounts to expressing the higher spin fields in the vielbein basis: $Y_{a_1\dots a_s}=e_{a_1}^{\mu_1}\dots e_{a_s}^{\mu_s}Y_{\mu_1\dots\mu_s}$ with $e_{a_i}^{\mu_i}=\frac{z}{R_{AdS}}\delta_{a_i}^{\mu_i}$.}
\begin{equation}
  Y_{\mu_1\dots \mu_s}=z^s\Phi_{\mu_1\dots\mu_s}
  \label{}
\end{equation}
as a multiplet of scalars. And the equation for $Y_{\mu_1\dots\mu_s}$ is
\begin{equation}
  \partial_\alpha\partial^\alpha Y_{\mu_1\dots\mu_s}+z^{d-1}\partial_z\left( z^{1-d}\partial_z Y_{\mu_1\dots\mu_s} \right)-\frac{(s-2)(s+d-2)}{z^2}Y_{\mu_1\dots\mu_s}=0
  \label{}
\end{equation}
which is just the free scalar equations with a mass parameter
\begin{equation}
  m^2=(s-2)(s+d-2)
  \label{}
\end{equation}
corresponding to scaling dimension
\begin{equation}
  \Delta=s+d-2
  \label{}
\end{equation}
The near-boundary behavior of $Y_{\mu_1\dots\mu_s}$ is
\begin{equation}
  Y_{\mu_1\dots\mu_s}\sim z^\Delta\mathcal{O}_{\mu_1\dots\mu_s}
  \label{}
\end{equation}
So one can directly construct the bulk spin- $s$ field:
\begin{equation}\label{hssmearing}
  \Phi_{\mu_1\dots\mu_s}=\frac{\Gamma\left( s+\frac{d}{2}-1 \right)}{\pi^{\frac{d}{2}}\Gamma\left( s-1\right)}\frac{1}{z^s}\int_{t'^2+|\textbf{y}'|^2<z^2}dt'd^{d-1}\textbf{y}'\left( \frac{z^2-t'^2-|\textbf{y}'|^2}{z} \right)^{s-2}\mathcal{O}_{\mu_1\dots\mu_s}(t+t',\textbf{x}+i\textbf{y}')
\end{equation}
for field with integer spin $s>1$. We see the field behaves like $z^{\Delta-s}=z^{d-2}$ near the boundary. Also, it gives rise to the expected scaling dimension of the conserved boundary currents.

%%%%%%%%%%%%%%%%%%%
\section{AdS covariance}\label{sec:4}
%%%%%%%%%%%%%%%%%%%

In this section we check the covariance of the construction. We apply Anti de Sitter group transformations to the local operator constructed in last section, and see if it is covariant up to gauge transformations.

The covariance under dilation is pretty straightforward, since both sides of (\ref{hssmearing}) have the same scaling dimension and thus rescaled in the same way when applied by a dilation. The special conformal transformations are less-trivial. The bulk AdS isometries that correspond to special conformal transformations are
\begin{align}
  & x^\mu\rightarrow x^\mu+2b\cdot x x^\mu-b^\mu\left( x^2+z^2 \right)\\
  & z\rightarrow z+2b\cdot x z
\end{align}
Also, for a higher spin field we've
\begin{equation}
  \Phi_{M_1\dots M_s}\rightarrow \Phi_{M_1\dots M_s}'=\frac{\partial x^{N_1} }{\partial x^{M_1}}\dots\frac{\partial x^{N_s}}{\partial x^{M_s}}\Phi_{N_1\dots N_s}
  \label{}
\end{equation}
which gives transformation laws for components:
\begin{align}
  & \delta\Phi_{zz\dots z}=\delta\Phi_{zz\dots \mu_s}=\dots=\delta\Phi_{zz\mu_3\dots\mu_s}=0\\
  & \delta\Phi_{z\mu_2\dots\mu_s}=2z b^{\alpha}\Phi_{\alpha\mu_2\dots\mu_s}\\
  & \delta\Phi_{\mu_1\dots\mu_s}=2b^\alpha\sum^s_{j=1} x_{\mu_j}\Phi_{\alpha\mu_1\dots\mu_{j-1}\mu_{j+1}\dots\mu_s}-2x^\alpha\sum^s_{j=1}b_{\mu_j}\Phi_{\alpha\mu_1\dots\mu_{j-1}\mu_{j+1}\dots\mu_s}-2s\left( b\cdot x \right)\Phi_{\mu_1\dots\mu_s}
\end{align}
%Modified the last.
These transformations bring the gauge field out of holographic gauge, which requires a compensating gauge transformation to recover the holographic gauge. Such a gauge transformation takes the form
\begin{align}
  & \delta\Phi_{z\mu_2\dots\mu_s}=\frac{1}{z^{2s-2}}\partial_z\epsilon_{\mu_2\dots\mu_s}\\
  & \delta\Phi_{\mu_1\dots\mu_s}=\frac{1}{z^{2s-2}}\sum_{j=1}^s\partial_{\mu_j}\epsilon_{\mu_1\dots\mu_{j-1}\mu_{j+1}\dots\mu_s}
\end{align}
with the gauge parameters
\begin{equation}
  \epsilon_{\mu_2\dots\mu_s}=-\frac{\Gamma(s+\frac{d}{2}-1)}{2^{2-s}\pi^{\frac{d}{2}}2\Gamma(s)}\int d^{d}x'\Theta(\sigma z')(\sigma zz')^{s-1} 2b^\alpha\mathcal{O}_{\alpha\mu_2\dots\mu_s}
  \label{}
\end{equation}
where we defined the AdS invariant length as
\begin{equation}
  \sigma=\frac{z^2+z'^2+(x-x')^2}{2zz'}
  \label{}
\end{equation}
and in the gauge parameter above, $\sigma z'$ and $\sigma zz'$ should be understood as in the limit $z'\rightarrow 0$ (they are `regularized' AdS invariant lengths). Then we have 
\begin{equation}
  \delta\Phi_{z\mu_2\dots \mu_s}=-2zb^\alpha\Phi_{\alpha\mu_2\dots\mu_s}
  \label{}
\end{equation}
and
%fixed
\begin{equation}
  \delta\Phi_{\mu_1\dots\mu_s}=-\frac{\Gamma(s+\frac{d}{2}-1)}{2^{2-s}\pi^{\frac{d}{2}}\Gamma(s-1)}\frac{1}{z^{2s-2}}\int d^dx\Theta(\sigma z')(\sigma zz')^{s-2}\sum_{j=1}^s(x-x')_{\mu_j}2b^\alpha\mathcal{O}_{\alpha\mu_1\dots\mu_{j-1}\mu_{j+1}\dots\mu_s}
  \label{}
\end{equation}
We see the $\Phi_{z\mu_2\dots\mu_s}$ components are brought back to zero and holographic gauge restored, while $\Phi_{\mu_1\dots\mu_s}$ components get an extra piece from gauge transformation which combines with the AdS transformation and gives
\begin{equation}
  z^s\delta\Phi_{\mu_1\dots\mu_s}=\frac{\Gamma(s+\frac{d}{2}-1)}{\pi^{\frac{d}{2}}\Gamma(s-1)}\int d^dx\Theta(\sigma z')(2\sigma z')^{s-2}\sum^s_{j=1}\left( 2x_{\mu_j}'b^\alpha\mathcal{O}_{\alpha\dots\mu_s}-2x^\alpha b_{\mu_j}\mathcal{O}_{\alpha\dots\mu_s} \right)-2s\left( b\cdot x \right)z^s\Phi_{\mu_1\dots\mu_s}
  \label{field_transf}
\end{equation}
This can be further simplified if we consider current conservation, which gives
\begin{equation}
  \int d^dx'\Theta(\sigma z')(\sigma zz')^{s-1}\partial^\alpha\Phi_{\alpha\mu_2\dots\mu_s}=0
  \label{}
\end{equation}
Integrating by parts, we get
\begin{equation}
  \int d^dx \Theta(\sigma z')\left( \sigma z' \right)^{s-2}(x-x')^\alpha\mathcal{O}_{\alpha\dots\mu_s}=0
  \label{}
\end{equation}
This suggests we can replace the $x'$ with $x$ in (\ref{field_transf}) above:
\begin{align}
  z^s\delta\Phi_{\mu_1\dots\mu_s}&=\frac{\Gamma(s+\frac{d}{2}-1)}{\pi^{\frac{d}{2}}\Gamma(s-1)}\int d^dx\Theta(\sigma z')(2\sigma z')^{s-2}\sum^s_{j=1}\left( 2x_{\mu_j}b^\alpha\mathcal{O}_{\alpha\dots\mu_s}-2x^\alpha b_{\mu_j}\mathcal{O}_{\alpha\dots\mu_s} \right)-2s\left( b\cdot x \right)z^s\Phi_{\mu_1\dots\mu_s}\\
  &=z^s\sum^s_{j=1}\left( 2x_{\mu_j}b^\alpha\Phi_{\alpha\dots\mu_s}-2x^\alpha b_{\mu_j}\Phi_{\alpha\dots\mu_s} \right)-2sz^s \left( b\cdot x \right)\Phi_{\mu_1\dots\mu_s}
  \label{}
\end{align}
which matches with the transformation of the boundary current under special conformal transformations when one combines it with the transformation of the integration measure:
\begin{equation}
  \delta\mathcal{O}_{\mu_1\dots\mu_s}=\sum^s_{j=1}\left( 2x_{\mu_j}b^\alpha\mathcal{O}_{\alpha\dots\mu_s}-2x^\alpha b_{\mu_j}\mathcal{O}_{\alpha\dots\mu_s} \right)-2(d+s-2) (b\cdot x) \mathcal{O}_{\mu_1\dots\mu_s}
  \label{}
\end{equation}

%%%%%%%%%%%%%%%%%%%%%%%
\section{Non-locality at two-point functions}\label{2pfn}
%%%%%%%%%%%%%%%%%%%%%%%

In this section we look at two-point functions between a spin- $s$ bulk gauge field and spin- $s$ boundary conserved current. As spin goes up, the structure of such correlation functions involve more and more complicated tensor structures. Hence a more convenient way of studying them is to contract the indices with null vectors, forming contracted operators without indices.

%%%%%%%%%%%%%%%%%%%%%%%%%%%%%
\subsection{Two-point function of operators contracted by same null vector}
%%%%%%%%%%%%%%%%%%%%%%%%%%%%%
Here we start from two-point function of spin- $s$ operators that are formed by contracting with a certain null vector $n^\mu$ for all the indices of both the spin- $s$ currents. The advantage of this approach is that we can keep the calculation under control as $s$ goes up, without worrying about the tensor structure which becomes more and more complicated otherwise. We lose some of the terms in the two-point function, because any term in the contraction that is proportional to $\eta^{\mu\nu}$ vanishes when contracted with $n_\mu n_\nu$. Still the correlation functions of these contracted tensors reflect generic features of the terms that appear in the full tensor structure. We will see that the two-point correlation function between a contracted bulk gauge field and a contracted boundary current exhibit both bulk and boundary lightcone singularities, and is thus acausal. This kind of acausality is allowed as it is due to the gauge redundancy of the fields we are constructing, and thus not causing any physical trouble.

The contracted spin- $s$ current is defined as:
\begin{equation}
  \mathcal{O}^{(s)}\equiv n^{\mu_1}\dots n^{\mu_s}\mathcal{O}_{\mu_1\dots\mu_s}~,~n^\mu n_\mu=0
  \label{}
\end{equation}
In a $d$-dimensional CFT, the two point function for such spin- $s$ conserved current operators is
\begin{equation}
  <\mathcal{O}^{(s)}(x)\mathcal{O}^{(s')}(0)>=\delta_{ss'}\left( \frac{1}{x^2} \right)^{s+d-2}\left( \frac{2(x\cdot n)^2}{x^2} \right)^s
  \label{}
\end{equation}
Then the two-point function of a spin- $s$ gauge field and a spin- $s$ conserved current is written as
\begin{align}
< \Phi^{(s)}(t,z,\textbf{x})\mathcal{O}^{(s)}(0)>=\frac{\Gamma\left( s+\frac{d}{2}-1 \right)}{\pi^{\frac{d}{2}}\Gamma\left( s-1\right)}\frac{1}{z^s}&\int_{t'^2+|\textbf{y}'|^2<z^2}dt 'd^{d-1}\textbf{y}'\left( \frac{z^2-t'^2-|\textbf{y}'|^2}{z} \right)^{s-2}\nonumber\\
&<\mathcal{O}^{(s)}(t+t',\textbf{x}+i\textbf{y}')\mathcal{O}^{(s)}(0)>
  \label{}
\end{align}
To evaluate this integral, we must handle the hidden powers of $x^{\mu_i}$. They give tensor structures that can be quite complicated for high spins. For example for spin-one we have
\begin{equation}
  \frac{x^\mu x^\nu}{(x^2)^d}=\frac{1}{2(d-1)}\eta^{\mu\nu}\left( \frac{1}{x^2} \right)^{d-1}+\frac{1}{4(d-1)(d-2)}\partial^\mu\partial^\nu\left( \frac{1}{x^2} \right)^{d-2}
  \label{}
\end{equation}

We notice that there is a term proportional to $\eta^{\mu\nu}$ which is eliminated when contracting with null vector $n_\mu$ and only the second term, which is a total derivative, remains. Generically for higher spin two-point functions, the tensor structure can always be written down as a sum of terms with different powers of $\eta_{\mu\nu}$ and derivatives. Then any term with $\eta_{\mu\nu}$ vanishes when contracting with null vectors and what remains is always a term with $2s$ derivatives on a $d$-dimensional scalar function:
\begin{equation}
  <\mathcal{O}^{(s)}(x)\mathcal{O}^{(s)}(0)>=2^s\frac{(x\cdot n)^{2s}}{(x^2)^{2s+d-2}}=\frac{(d-3)!}{2^{s}(d+2s-3)!}(n\cdot \partial)^{2s}\left( \frac{1}{x^2} \right)^{d-2}
  \label{}
\end{equation}
Thus the two-point function here is obtained by applying derivatives on a certain integral:
\begin{equation}
  < \Phi^{(s)}(t,z,\textbf{x})\mathcal{O}^{(s)}(0)>=\frac{\Gamma\left( s+\frac{d}{2}-1 \right)}{\pi^{\frac{d}{2}}\Gamma\left( s-1\right)}\frac{(d-3)!}{2^{s}(d+2s-3)!}\frac{1}{z^s}\left( n\cdot\partial \right)^{2s}J(s,d)
  \label{}
\end{equation}
with 
\begin{equation}
  J(s,d)\equiv \int_{t'^2+|\textbf{y}'|^2<z^2}dt' d^{d-1}\textbf{y}'\left( \frac{z^2-t'^2-|\textbf{y}'|^2}{z} \right)^{s-2}\left( \frac{1}{-(t+t')^2+|\textbf{x}+i\textbf{y}'|^2} \right)^{d-2}
  \label{}
\end{equation}
The integral is evaluated to be
\begin{equation}\label{jsd}
J(s,d)=\frac{\pi^{\frac{d}{2}}\Gamma(s-1)}{\Gamma\left( s+\frac{d}{2}-1 \right)}\frac{z^{s+d-2}}{x^{2(d-2)}}F\left( d-2,\frac{d}{2}-1,s+\frac{d}{2}-1,-\frac{z^2}{x^2} \right)
\end{equation}
We see that in general it gets singularities on boundary lightcone from the factor $\frac{1}{x^{2(d-2)}}$, as well as on bulk lightcone from the hypergeometric function. The bulk singularity is what one would expect because the bulk locality demands the operators to commute at space-like separation and fail to commute at time-like separation. Thus we need a singularity on the lightcone as otherwise the property of operators being commuting at space-like separation can be analytically continued outside the spatial lightcone without any obstacle, thus failing to give a non-vanishing commutator at time-like separation and vice-versa. But the boundary lightcone doesn't necessarily imply the same from the bulk point of view; singularities on the boundary lightcone means that two bulk operators fail to commute even when they are space-like separated, as long as they appear to be time-like separated when projected onto the boundary. Therefore the two-point functions appear to be non-local. But since they are gauge-dependent, such acausality is allowed.

%%%%%%%%%%%%%%%%%%%%%%%%%
\subsection{Correlation function with different null vectors}
%%%%%%%%%%%%%%%%%%%%%%%%%

In the previous subsection, we have computed the two-point function between two spin- $s$ boundary operators contracted with the same null vector. As we have mentioned before, doing so makes us lose some information about the two-point function and is usually not how they are generally treated in higher spin theories. Here we generalize the calculation by introducing different null vectors \cite{Stanev:2012nq} separately for the two spin- $s$ operators. The final bulk-boundary propagator is then obtained as a sum of terms with each term showing both the bulk and boundary lightcone singularities.

The conformal invariant two-point function for the boundary currents is
\begin{equation}
\langle \mathcal{O}_{s_1}(n,x_1)\mathcal{O}_{s_2}(n',x_2)\rangle=\delta_{s_1s_2}c(s_1)R_{nn'}^{s_1}\left(\frac{1}{x_{12}^2}\right)^{d-2}
\end{equation}
Here $c(s_1)$ is the normalization of two-point function in the CFT which we can set to 1 for all the gauge fields. On the other hand, 
\[
R_{nn'}=\frac{1}{x_{12}^2}\left(n\cdot n'-\frac{2(n\cdot x_{12})(n'\cdot x_{12})}{x_{12}^2}\right)
\]
This quantity encodes part of the full tensor structure one gets in the full correlation functions. Here we are contracting all the indices of an operator with a certain null vector, and for the operators in the correlation function we assign an independent null vector for each. That is, in this case we can have different null vectors for the first and second operators in the two-point function. Notice that once we set $n=n'$ in the expression above we are back to the two-point function computed in the section above. Thus it can be thought of as a generalization of the calculation there.

Continuing with different null vectors, the two-point function can be decomposed into a sum using binomial theorem:
\begin{equation}
\langle \mathcal{O}_{s_1}(n,x_1)\mathcal{O}_{s_2}(n',x_2)\rangle=\sum_{m}\delta_{s_1s_2}c(s_1,m)(n\cdot n')^{s_1-m}\left(\frac{1}{x_{12}^2}\right)^{d+s_1-2}\left(\frac{2(n\cdot x_{12})(n'\cdot x_{12})}{x_{12}^2}\right)^m
\end{equation}

Now we smear the first operator in the two-point function with the prescription established in the previous sections:
\begin{eqnarray}
z^{s}\langle \Phi^{(s)}(n,x_1)\mathcal{O}_{s}(n',x_2)\rangle=\frac{\Gamma\left(s+\frac{d}{2}-1\right)}{\Gamma\left(s-1\right)\pi^{d/2}}\int_{B^d}dt'd^{d-1}\textbf{y}'\left(\sigma z'\right)^{\Delta-d}\nonumber\\
\sum_{m}c(s,m)(n\cdot n')^{s-m}\frac{1}{x_{12}^{d+s-2}}\left(\frac{2(n\cdot x_{12})(n'\cdot x_{12})}{x_{12}^2}\right)^m
\end{eqnarray}
where the integration is over a ball of radius $z$ (and $z'\to 0$ for normalization). The only combination for the terms in the parenthesis, that transform covariantly under conformal transformation, is given by a combination of $\eta_{\mu\nu}$ and partial derivatives (for an analogous expression for spin-1 and 2 cases, see \cite{KLRS}) and the factors of $\frac{1}{x_{12}^2}$ associated with it could be found out by dimensional analysis. These facts give a unique series expansion for the bulk-boundary correlator given by\footnote{Of course the spatial direction of $x_{1}$ in the integrand should be thought of as having analytic continuation to imaginary $y_1$ direction.}
\begin{eqnarray}
z^{s}\langle \Phi^{(s)}(n,x_1)\mathcal{O}_{s}(n',x_2)\rangle=\frac{\Gamma\left(s+\frac{d}{2}-1\right)}{\Gamma\left(s-1\right)\pi^{d/2}}\sum_{m}c(s)\int_{B^d}dt'd^{d-1}\textbf{y}'\left(\sigma z'\right)^{\Delta-d}\nonumber\\
(n\cdot n')^{s-m}\left(c_1(\eta_{\mu\nu})^m\left(\frac{1}{x_{12}^2}\right)^{d+s-2}+\dots+c_m\left(\partial_\mu\partial_\nu\dots(\mbox{$m$ terms})\right)\left(\frac{1}{x_{12}^2}\right)^{d+s-m-2}\right)
\end{eqnarray}
Now to understand the causal structure of the field $\Phi^{(s)}(n,x)$, we define, as in \cite{KLRS}, the quantity
\begin{equation}\label{Jdefn}
J_m(s,d)=\int_{B^d}dt'd^{d-1}\textbf{y}'\left(\sigma z'\right)^{\Delta-d}\left(\frac{1}{x_{12}^2}\right)^{d-m}
\end{equation}
So we need to understand the causal structures of $J_{2-s}(s,d)$ for various values of $s$ and their derivatives. Once again following the same steps as in \cite{KLRS} or the last subsection, we find that they contain both bulk and boundary light cone singularities via terms like (taking $x_2=0$ e.g. and calling $x_1=x$)
\[
J_{2-s}(s,d)\sim\frac{z^{s+d-2}}{x^{2(d-2+s)}}F\left(d-2+s,b,b,-\frac{z^2}{x^2}\right),\quad b=\frac{d}{2}-1+s
\]
Because each of the terms in the bulk-boundary correlator has both the bulk and boundary lightcone singularities, once again the gauge fields $\Phi^{(s)}(n,x)$ don't satisfy bulk locality.

%%%%%%%%%%%%%%%%%%%%%%%%%
\section{Non-locality at three-point functions}\label{sec:6}
%%%%%%%%%%%%%%%%%%%%%%%%%

To understand the dynamics of quantum gravity in the presence of higher spins in AdS space, the first step will be to go beyond free bulk theory and to introduce interactions between the higher spins, i.e. to do $1/N$ perturbation. Although as $N$ is still large, we can expect the bulk theory to have properties that we will expect in semiclassical theories. Indeed in \cite{KLL}, \cite{KL} and \cite{Kabat:2013wga} it was shown using the smearing technique that up to spin-2, it is possible to construct local bulk operators just from the principles of  microcausality. In this section we study the same for spins higher than 2. More precisely we study the three point function involving a higher spin bulk gauge field and boundary higher spin conserved currents (using the leading order smearing function)\footnote{One may ask whether the leading order (in $1/N$ expansion) smearing function is the correct smearing function to use in a three-point function. It is not. In fact, this is a way to see that such smearing function is insufficient and to avoid the singularity structure, probably one needs to add higher dimensional multitrace operators. However, as these are massless gauge fields, one will find out (like the cases for spin-1 and spin-2 as in \cite{KL},\cite{Kabat:2013wga}) that such modifications are still insufficient to make the gauge fields local. We don't show it here.}. We find that once again the gauge fields fail to satisfy bulk causality. For technical reasons, our results are more limited from the earlier results up to spin 2. For example, we do not find out the precise `gauge invariant' higher spin boundary currents which, if exist, should have the correct locality properties with e.g. a charged bulk scalar (whose definition might involve the need to include a smeared tower of non-primary boundary operators\footnote{The appearance of non-primary operators turn out to be necessary as the charged bulk scalars transform anomalously under AdS isometries.} \cite{KL}). Also, e.g. we do not find the precise bulk gauge invariant higher spin operator, which, presumably upon modifications by smearing of higher (conformal) dimensional boundary currents, will commute with another boundary current.\footnote{However the statement of gauge invariance from three-point function level and from spin-2 case is already quite uncertain. For example, the field strength tensor for spin-1 remains a local operator at order $1/N$. But, it is not clear whether the same is true for Weyl tensor \cite{Kabat:2013wga}.} However, we refer the readers to section \ref{sec:conclusion} where we briefly comment on bulk locality at higher orders in $\frac{1}{N}$.

The three-point function for higher spin conserved current in CFT$_d$ has been given by \cite{Zhiboedov:2012bm} e.g.\footnote{Equivalent expressions are given in e.g. \cite{Costa:2011mg}. But their construction is based on the embedding formalism of AdS space (discussed therein). The smearing construction in embedding language is briefly discussed in \cite{Xiao:2014uea}. See also \cite{GY}, \cite{Stanev:2012nq}.} It takes the form
\[
\langle \mathcal{O}_{s_1}(x_1)\mathcal{O}_{s_2}(x_2)\mathcal{O}_{s_3}(x_3)\rangle=\frac{\langle\langle \mathcal{O}_{s_1}\mathcal{O}_{s_2}\mathcal{O}_{s_3}\rangle\rangle}{|x_{12}x_{13}x_{23}|^{d-2}}
\]
Once again different currents have been contracted by different null vectors to keep the treatment general. Here
\[
\langle\langle \mathcal{O}_{s_1}\mathcal{O}_{s_2}\mathcal{O}_{s_3}\rangle\rangle=\sum_{k=0}^{min[s_1,s_2,s_3]/2}c_k\langle\langle \mathcal{O}_{s_1}\mathcal{O}_{s_2}\mathcal{O}_{s_3}\rangle\rangle^k
\]
with 
\begin{align}\label{3p}
&\langle\langle \mathcal{O}_{s_1}\mathcal{O}_{s_2}\mathcal{O}_{s_3}\rangle\rangle^k=\prod_{i=1}^{3}e^{V_i}\prod_{i<j=1}^{3}{}_0F_1\left(\frac{d}{2}+2k-1,-\frac{R_{ij}}{2}\right)\Lambda_1^{2k}F\left(\frac{1}{2}-k,-k,3-\frac{d}{2}-2k,-\frac{\Lambda_2}{2\Lambda_1^2}\right)\nonumber\\
&\Lambda_1=V_1V_2V_3+\frac{1}{2}(V_1R_{23}+V_2R_{13}+V_3R_{12}), \qquad \Lambda_2=R_{12}R_{13}R_{23}\nonumber\\
&V_i=\frac{a_ix_{i,i+2}}{x_{i,i+2}^2}-\frac{a_ix_{i,i+1}}{x_{i,i+1}^2} \quad\mbox{and}\quad R_{ij}=\frac{1}{x_{ij}^2}\left(a_ia_j-2\frac{(a_ix_{ij})(a_jx_{ij})}{x_{ij}^2}\right)
\end{align}
The only difference here is that the null vectors $a$'s have dimensions of length, unlike $n$'s introduced in the previous section. This means the index-contracted conserved currents have dimension $d-2$. At the end to get three point correlator, we should pick up the term proportional to $a_1^{s_1}a_2^{s_2}a_3^{s_3}$. Under a conformal transformation $x\to \hat{x}$, these null polarization tensors also transform as 
\[
a^\mu_i\to\hat{a}^\mu_i=\frac{\partial \hat{x}^\mu}{\partial x^\nu}\Bigg|_{x=x_i} a_i^\nu
\] 
%At the light cone limit $\langle\langle \mathcal{O}_{s_1}\mathcal{O}_{s_2}\mathcal{O}_{s_3}\rangle\rangle^k\sim(x^+)^{2k}$ and that the spins have been understood in that sense.
All one needs to do now is to smear the first operator $\mathcal{O}_{s_1}$ to construct a higher spin bulk operator using the higher spin smearing function and study its commutator with another operator (e.g. $\mathcal{O}_{s_2}$) where they are spacelike separated from one another. Once again, to realize that the commutator changes as we go from a timelike to spacelike separation, it suffices to consider a bulk-boundary three-point function and study its singularity structures:
\begin{align}\label{3pfn}
&< \Phi^{(s)}(t_1,z,\textbf{x}_1)\mathcal{O}^{(s)}(0)\mathcal{O}^{(s)}(x_3)>\nonumber\\
&=\frac{\Gamma\left( s+\frac{d}{2}-1 \right)}{\pi^{\frac{d}{2}}\Gamma\left( s-1\right)}\frac{1}{z^s}\int_{t_1'^2+|\textbf{y}_1'|^2<z^2}dt_1 'd^{d-1}\textbf{y}_1'\left( \frac{z^2-t_1'^2-|\textbf{y}_1'|^2}{z} \right)^{s-2}<\mathcal{O}^{(s)}(t_1+t_1',\textbf{x}_1+i\textbf{y}_1')\mathcal{O}^{(s)}(0)\mathcal{O}^{(s)}(x_3)>
\end{align}
Once again we expect the answer to involve both bulk and boundary lightcone singularities. But as is clear from (\ref{3p}), the full calculation is quite messy. Below we will make some approximations to simplify the calculation and to clearly see the structure of the non-locality.

We start with $x_2=0$ and $x_3\to$ large. Also, as we are most interested in the $x_1$ integral, we will mainly concentrate on the $x_1$ dependent terms below\footnote{Note that at the end all the terms are multiplied together, so it's okay to work this way.}. We see that only $V_1$ and $V_2$ has $x_1$ dependence which go as $-\frac{a_1}{x_1}-\frac{a_1}{x_3}$ and $-\frac{a_2}{x_1}+\frac{a_2}{x_3}$ respectively. On the other hand $V_3\to 0$. Similarly only $R_{12}$ has $x_1$ dependence which goes as $\frac{1}{x_1^2}$. As a result, $\Lambda_1$ has no $x_1$ dependence and $\Lambda_2\to \frac{1}{x_1^2}$. We will also need to see how the hypergeometric functions ${}_0F_1$ and $F$ behave. Using their series expansions, we see that in this limit
\[
{}_0F_1\left(\frac{d}{2}+2k-1,-\frac{R_{ij}}{2}\right), F\left(\frac{1}{2}-k,-k,3-\frac{d}{2}-2k,-\frac{\Lambda_2}{2\Lambda_1^2}\right)\sim \sum_{m=0}^{\infty}\left(\frac{1}{x_1^2}\right)^m
\]
So, at the end the integral at the right hand side of (\ref{3pfn}) contains terms like
\begin{align}
J'(s,d)&\equiv \sum_{m,n=0}^{\infty}\int_{t_1'^2+|\textbf{y}_1'|^2<z^2}dt_1' d^{d-1}\textbf{y}_1'\left( \frac{z^2-t_1'^2-|\textbf{y}_1'|^2}{z} \right)^{s-2}\nonumber\\
&\left( \frac{1}{-(t_1+t_1')^2+|\textbf{x}_1+i\textbf{y}_1'|^2} \right)^{\frac{d-2}{2}+m+n}e^{-(a_1+a_2)\left(\frac{1}{-(t_1+t_1')^2+|\textbf{x}_1+i\textbf{y}_1'|^2}\right)^{\frac{1}{2}}}\nonumber\\
&\sim \sum_{m,n,l=0}^{\infty}\int_{t_1'^2+|\textbf{y}_1'|^2<z^2}dt_1' d^{d-1}\textbf{y}_1'\left( \frac{z^2-t_1'^2-|\textbf{y}_1'|^2}{z} \right)^{s-2}
\left( \frac{1}{-(t_1+t_1')^2+|\textbf{x}_1+i\textbf{y}_1'|^2} \right)^{\frac{d-2}{2}+m+n+\frac{l}{2}}
\end{align}
where in the last line we have expanded the exponential. We see that each term of the series is just like (\ref{Jdefn}). In fact we have
\[
J'(s,d)\sim\sum_{m,n,l}\frac{\pi^{\frac{d}{2}}\Gamma(s-1)}{\Gamma\left(s+\frac{d}{2}-1\right)}\frac{z^{s+d-2}}{x^{2\left(\frac{d-2}{2}+m+n+\frac{l}{2}\right)}}F\left(\frac{d-2}{2}+m+n+\frac{l}{2},m+n+\frac{l}{2},s+\frac{d}{2}-1,-\frac{z^2}{x^2}\right)
\]
For each term in the series we have both boundary lightcone singularities as $x^2\to 0$ and bulk lightcone singularities via the $\frac{z^2}{x^2}$ factor in the hypergeometric function. They survive for every $m,n,l\in\mathbb{Z}^+$. We should however note various shortcomings of the above calculation. Not only that we don't have a closed form expression, the result is also obtained at particular limiting values of the boundary coordinates and is hence non-covariant. It will be fulfilling to obtain a result in fully covariant form in terms of AdS covariant distances of various operators.

%%%%%%%%%%%%%%%%%%%%%%%%%%
%\section{Bulk Locality for Higher Spin Fields}\label{sec:7}
%%%%%%%%%%%%%%%%%%%%%%%%%%

%%%%%%%%%%%%%%%%%%%%%%%%%
\section{Higher spins in de Sitter}\label{hs_ds}
%%%%%%%%%%%%%%%%%%%%%%%%%

In this section we change the gear slightly and discuss the higher spin construction in dS space. As shown in \cite{Xiao:2014uea}, a local scalar field in dS is given by smearing two sets of boundary operators of `complementary' conformal dimensions $\mathcal{O}_{\Delta}$ and $\mathcal{O}_{d-\Delta}$. For example, for bulk operators with $m^2>\left(\frac{d}{2}\right)^2$ one has ($\eta$ here is the time coordinate; see (\ref{dSmetric}) below)
\begin{eqnarray}
\Phi(\eta,\mathbf{x})=\frac{\Gamma\left(\Delta-\frac{d}{2}+1\right)}{\pi^{d/2}\Gamma\left(\Delta-d+1\right)}\int_{|x'|<\eta}d^d\mathbf{x'}\left(\frac{\eta^2-\mathbf{x'}^2}{\eta}\right)^{\Delta-d}\mathcal{O}_{\Delta}(\mathbf{x}+\mathbf{x'})\nonumber\\+\frac{\Gamma\left(\frac{d}{2}-\Delta+1\right)}{\pi^{d/2}\Gamma\left(1-\Delta\right)}\int_{|x'|<\eta}d^d\mathbf{x'}\left(\frac{\eta^2-\mathbf{x'}^2}{\eta}\right)^{-\Delta}\mathcal{O}_{d-\Delta}(\mathbf{x}+\mathbf{x'})
\end{eqnarray}
However, there are few subtle issues regarding constructing higher spin fields in de Sitter background, namely the higher spin fields in dS behave like scalars in dS with $m^2<(d/2)^2$. This is problematic as then their boundary components fall off at different rates and aren't oscillatory (although this is not a pathology itself, as the boundary theory is non-unitary). Thus they don't always have a good positive/ negative frequency interpretation (although, if the spins are non-integers or we have non-integer conformal dimensions, one can analytically continue the construction of $m^2>(d/2)^2$ in dS to this case). Hence to tackle the higher spins in dS, a minimalist point of view was taken in \cite{Xiao:2014uea}. Namely, one can just analytically continue the AdS result of higher spins to the dS case by an analytic continuation of coordinates. However, this construction is manifestly acausal as here one simply ignores either the positive or negative frequency modes of the bulk fields\footnote{Note that this is acausality of construction, not the acausal property of higher spin gauge fields discussed before. This way one can't even construct a higher spin gauge invariant operator which is causal.}. To make a causal construction in dS, one must incorporate both the boundary operators and sources \cite{Xiao:2014uea},\cite{Sarkar:2014jia}.

More technically, if one wants to use the dS scalar smearing function with $\Delta=s+d-2$, one finds that for $s>1$ ($s, d$ integers) there are divergences from the gamma functions (appearing as the pre-factors of the smearing function) and for $s=1$, there's a problem with splitting of Wightman function of Euclidean vacuum, which in turn makes it impossible to write the bulk field in terms of smearing two sets of boundary operators over a finite support.

However the technicalities could be overcome, if one starts from AdS smearing function on a cut-off surface (without imposing the normalizability condition) and then analytically continues to dS and take the boundary limit. In \cite{Sarkar:2014jia} it was done for massless scalars in AdS$_2$ and it reproduces the correct dS$_2$ massless scalar smearing (thereby all the causality properties of its correlator). It is also possible to extend it for higher dimensions and massive cases, although it is then technically difficult to confine the smearing support over a finite region. But nonetheless a construction is possible. Now to address the specific mass range in question, i.e. dS with $m^2<(d/2)^2$ one can simply start from higher spin fields on AdS (for which $m^2> -(d/2)^2$) at the cut-off surface and analytically continue the result to dS. Not only it should then reproduce the required smearing function, it also corresponds to the required mass range for dS. But as the behavior of higher spin fields in AdS is nothing but similar to massive scalars in AdS, one can simply borrow the results of scalar smearing on the cut-off surface.

To begin, we start with our results in section \ref{sec:3}, where we found out that the higher spin fields in AdS behave like massive scalars which correspond to boundary fields of conformal dimension $\Delta=s+d-2$. We assume that the cut-off surface is located at $z=z_0$ and as the standard near boundary behavior of a scalar field, it also behaves around the cut-off surface as 
\begin{eqnarray}\label{jphidefnco}
j_{cut}(x,z_0)=z^{-d+\Delta}\Phi(z,x)|_{z\to z_0}\qquad\mbox{and}\nonumber\\\phi_{b,cut}(x,z_0)=z^{-2\nu}z\partial_z(z^{-d+\Delta}\Phi)|_{z\to z_0}
\end{eqnarray}
%then we get
%\begin{equation}\label{reln}
%\phi_{b,cut}|_{z_0\to 0}=\phi_b\qquad\mbox{and}\qquad j_{cut}|_{z_0\to 0}=j+\frac{\phi_b}{2\nu}z_0^{2\nu}=j
%\end{equation}
Note that we get the correct $z\to 0$ behavior from here as we take $z_0\to 0$ and also that as we take $z_0\to 0$, the boundary value operators $\phi_{b,cut}$ and $j_{cut}$ become boundary operators of dimensions $\Delta$ and $d-\Delta$ respectively (as one needs for dS smearing functions). Note that usually these two operators are related and not independent in AdS construction, because of the imposition of the normalizability condition. But as our goal is to connect it with the dS result, we will not impose any such condition (which is once again compatible with the non-unitarity of the dS boundary theory). For a massive scalar $\Phi$, which corresponds to a boundary operator of dimension $\Delta$, standard AdS/ CFT dictates a free massive bulk scalar $\Phi$ has two independent solutions \cite{Balasubramanian:1998sn}. Their Fourier components are
\[
\Phi^{\pm}(z,x)=e^{-i\omega t+i\mathbf{k}\cdot\mathbf{x}}z^{d/2}J_{\pm\nu}(\sqrt{\omega^2-k^2}z)=e^{iqx}z^{d/2}J_{\pm\nu}(|q|z)
\] 
with\footnote{For $\nu\in\mathbb{Z}$, $J_{-\nu}$ is replaced by $Y_\nu$. For simplicity we can also assume $\nu>0$ which is compatible with $m^2> -(d/2)^2$ for AdS.}
\[
\nu=\Delta-\frac{d}{2}=\sqrt{\frac{d^2}{4}+m^2}, \quad q=(\omega,\mathbf{k})\quad\mbox{and}\quad q^2=(k^2-\omega^2)<0
\]
The cut-off surface smearing function then becomes \cite{Sarkar:2014jia}
\begin{equation}\label{k1k2_1}
\Phi(z,x)=\int d^dx'K_1(x'|x,z,z_0)\phi_{b,cut}(x',z_0)+\int d^dx'K_2(x'|x,z,z_0)j_{cut}(x',z_0)
\end{equation}
where 
\begin{align}
K_1&=&\int_{\omega>|k|} \frac{d^dq}{(2\pi)^d}e^{iq(x-x')}\frac{\pi z_0^{\nu} z^{\frac{d}{2}}\csc{\nu\pi}}{2}\left(-J_{\nu}(qz_0)J_{-\nu}(qz)+J_{-\nu}(qz_0)J_{\nu}(qz)\right)\nonumber\\
K_2&=&\int_{\omega>|k|} \frac{d^dq}{(2\pi)^d}e^{iq(x-x')}\frac{\pi qz_0^{1-\nu} z^{\frac{d}{2}}\csc{\nu\pi}}{2}\left(J_{1-\nu}(qz_0)J_{\nu}(qz)+J_{\nu-1}(qz_0)J_{-\nu}(qz)\right)
\end{align}
for non-integer $\nu$ and 
\begin{align}
K_1&=&\int_{\omega>|k|} \frac{d^dq}{(2\pi)^d}e^{iq(x-x')}z_0^{\nu-1} z^{\frac{d}{2}}\frac{\left(J_{\nu}(qz_0)Y_{\nu}(qz)-Y_{\nu}(qz_0)J_{\nu}(qz)\right)}{q\left(J_{1-\nu}(qz_0)J_\nu(qz_0)-Y_{\nu-1}(qz_0)Y_{\nu}(qz_0)\right)}\nonumber\\
K_2&=&\int_{\omega>|k|} \frac{d^dq}{(2\pi)^d}e^{iq(x-x')}z_0^{-\nu} z^{\frac{d}{2}}\frac{\left(J_{1-\nu}(qz_0)J_{\nu}(qz)-Y_{\nu-1}(qz_0)Y_{\nu}(qz)\right)}{J_{1-\nu}(qz_0)J_\nu(qz_0)-Y_{\nu-1}(qz_0)Y_{\nu}(qz_0)}
\end{align}
for integer values of $\nu$. We now use the analytic continuation to go from the AdS to dS patch. In (\ref{AdS_PP}), we put (also taking $R_{dS}=1$)
\begin{eqnarray}\label{dS2AdSancont}
z\to \eta,\quad t\to t, \quad x^i\to ix^i \quad \mbox{and} \quad R_{AdS}\to iR_{dS}
\end{eqnarray}
to obtain the dS metric as 
\begin{equation}\label{dSmetric}
ds^2_{dS}=\frac{-d\eta^2+d\mathbf{x}^2}{\eta^2}
\end{equation}
The dS smearing function would then be given by the $\eta_0\to 0$ limit of (now $\mathbf{q}$ being a $d$-dimensional vector conjugate to the `spatial' coordinates $\mathbf{x}$)\footnote{At this limit, the fields $j_{cut}$ and $\phi_{b,cut}$ should be interpreted as $\mathcal{O}_{d-\Delta}$ and $\mathcal{O}_\Delta$ respectively.}
\begin{align}
K_1&=&\int\frac{d^d\mathbf{q}}{(2\pi)^d}e^{i\mathbf{q}\cdot(\mathbf{x}-\mathbf{x'})}\frac{\pi \eta_0^{\nu} \eta^{\frac{d}{2}}\csc{\nu\pi}}{2}\left(-J_{\nu}(\mathbf{q}\eta_0)J_{-\nu}(\mathbf{q}\eta)+J_{-\nu}(\mathbf{q}\eta_0)J_{\nu}(\mathbf{q}\eta)\right)\nonumber\\
K_2&=&\int \frac{d^d\mathbf{q}}{(2\pi)^d}e^{i\mathbf{q}\cdot(\mathbf{x}-\mathbf{x'})}\frac{\pi \mathbf{q}\eta_0^{1-\nu} \eta^{\frac{d}{2}}\csc{\nu\pi}}{2}\left(J_{1-\nu}(\mathbf{q}\eta_0)J_{\nu}(\mathbf{q}\eta)+J_{\nu-1}(\mathbf{q}\eta_0)J_{-\nu}(\mathbf{q}\eta)\right)
\end{align}
for non-integer $\nu$ e.g. Unfortunately, doing these integrals analytically (and taking the limit of $\eta_0\to 0$) seem technically challenging. But they could be carried out example by example. The finite support of the smearing functions should also then be clear. As mentioned before, starting with the massless scalars in AdS$_2$ is simplest and using these techniques, it correctly reproduces the known dS$_2$ results with finite support. In that case, the result is
\begin{eqnarray}
\Phi(\eta,x)=\frac{1}{2}\left[\mathcal{O}_{d-\Delta}(x+\eta)+\mathcal{O}_{d-\Delta}(x-\eta)\right]+\frac{1}{2}\int_{x-\eta}^{x+\eta} dx'\mathcal{O}_\Delta(x')
\end{eqnarray}

%%%%%%%%%%%%%%%%%%%%%%%%%%%%
\section{Conclusion and outlooks}\label{sec:conclusion}
%%%%%%%%%%%%%%%%%%%%%%%%%%%%

Here the main result of our work is a complete field theoretic construction of higher spin bulk gauge fields in AdS and dS which also appear in various higher spin AdS/ CFT and dS/ CFT dualities. We found that the higher spin fields constructed this way is AdS covariant and from studying their two-point correlators it is evident that they have the non-local behavior, which one expects from a generalized Gauss law. However, there are also some very interesting questions that remain to be answered. Namely, it will be really satisfying to see that such non-localities are precisely what one expects from Gauss law type constraints and whether one can finally build out a local gauge invariant quantity (like a higher dimensional analog of Weyl tensor) completely out of the boundary observables.\footnote{One such possible candidate is the Freedman- de Wit tensor \cite{de Wit:1979pe}, or rather its traceless part.} The higher orders in $\frac{1}{N}$ are also particularly interesting for reasons mentioned in the introduction. Here We took the first step to study, the already complicated three-point bulk to boundary propagators and noticed that if one uses the leading order smearing function, it is still possible to distinguish the bulk and boundary singularities. Before pinpointing the correct smearing function for this order or to study higher point function, a natural direction seems to just pursue the scalar field construction for four-point function to study bulk locality \cite{toappear}.

As mentioned before, it is not possible to gauge away the boundary singularities for higher spin gauge fields because such non-locality is needed from the point of view of bulk Gauss laws \cite{Heem}, \cite{KLRS}, \cite{Heemskerk:2012mn}. But it is natural to anticipate locality for gauge-invariant tensors like $F_{\mu\nu}$ or Weyl tensors which also exist for higher spin theories.\footnote{More precisely we want Weyl like higher rank tensors which has a vanishing vacuum expectation value in empty AdS space and transform homogeneously under coordinate transformations.} For now we will leave such calculations involving `Weyl tensors' for future work. However here we point out the direction towards the answers to some of the previous questions.

For example, one can also try to do an alternative calculation to convince themselves that the non-locality that we get in (\ref{3pfn}) is at least expected by the bulk Gauss constraints. A possible direction will be to do a calculation similar to \cite{KL} where it was shown for a boundary gauge current that inside a three-point function with other bulk charged scalars (charged under a $U(1)$ gauge field), it satisfies the correct Gauss constraint. To this end, one can start with a three point function between a higher spin conserved current on the boundary and two boundary spin-0 operators with dimension $\Delta$ \cite{Giombi:2011rz} (as required by Ward identity). For simplicity here we stick to CFT$_3$ so that the current $\mathcal{O}_{\mu_1\dots\mu_s}$ has dimension $(s+1)$. With the help of (\ref{3p}), we get
\[
\langle 
\mathcal{O}_s(x_1,a_1)\mathcal{O}_\Delta(x_2)\bar{\mathcal{O}}_\Delta(x_3)\rangle=\frac{c}{|x_{12}||x_{23}|^{2\Delta-1}|x_{13}|}(2V_1)^s
\]
Here $\mathcal{O}_s(x_1,a_1)$ are defined as contracting boundary higher spin currents by dimensionless null vectors as in section \ref{2pfn} and $c$ is a structure constant.  Also here the boundary coordinates $x_1,x_2$ and $x_3$ have been taken to have equal time components for simplicity. We can now smear one of the scalars to the bulk to obtain\footnote{Once again $x_2$'s appearing on the right hand side of the equation below is to be understood as already analytically continued to $y'$ in their spatial directions.}
\begin{eqnarray}\label{gauss3p}
\langle 
\mathcal{O}_s(x_1,a_1)\Phi(x_2,z)\bar{\mathcal{O}}_\Delta(x_3)\rangle=\frac{c\Gamma\left(\Delta-\frac{1}{2}\right)}{\pi^{3/2}\Gamma(\Delta-2)|x_{13}|}\int_{t'^2+|\textbf{y}'|^2<z^2}dt' d^{2}\mathbf{y}'\left(\frac{z^2-t'^2-|\textbf{y}'|^2}{z}\right)^{\Delta-3}\nonumber\\
\frac{\left[2\left(\frac{a_1x_{13}}{x_{13}^2}-\frac{a_1x_{12}}{x_{12}^2}\right)\right]^s}{|x_{12}||x_{23}|^{2\Delta-1}}
\end{eqnarray}
The integration appearing here is completely similar to (\ref{Jdefn}). Hence we note that the boundary gauge invariant higher spin current doesn't commute with bulk scalars also as the integration inevitably gives both bulk and boundary lightcone singularities as $x_{12}\to 0$ (this is of course extendable to higher dimensions). 

But to show the compatibility of this non-locality with the bulk Gauss constraint, the next step will be to calculate the commutator $\langle[\mathcal{O}_s(x_1,a_1),\Phi(x_2,z)]\bar{\mathcal{O}}_\Delta(x_3)\rangle$ where now we smear the boundary operator of dimension $\Delta$ to get a boundary scalar field $\Phi$. Here according to Gauss law, if we integrate $\mathcal{O}_s(x_1,a_1)$ over the two boundary spatial coordinates $\mathbf{x_1}$, one expects to obtain\footnote{Of course even then it won't be a full proof, but only a first order check. For a complete proof one needs to show that such relations are obtained as we add smeared higher dimensional operators to $\Phi$ \cite{KL}.} \cite{KLL}, \cite{Maldacena:2011jn}
\begin{equation}
-\langle\partial^{s-1}\Phi(x_2,z)\bar{\mathcal{O}}_\Delta(x_3)\rangle\sim\partial^{s-1}\left(\frac{z}{z^2+x_{23}^2}\right)^{\Delta}
\end{equation}
For example, the fact that we can expect terms like $z^\Delta$ by integrating (\ref{gauss3p}) is clear from the expression of $J_n(s,d)$ functions (\ref{Jdefn}). Similarly, one other possible check at the level of three point function will be to consider two bulk scalar fields charged under the higher spin currents and to check whether they commute between each other. 

However aside from these technicalities, there is a bigger goal to all this. One of the major outcome of current research is the understanding of higher spins as some limits of string theory itself \cite{Chang:2012kt}, \cite{Gaberdiel:2014cha}. And at least for large central charges we now have a complete identification of higher spins from the corresponding boundary operators which is completely well defined. Thus it remains to be seen if our present work can play a key role for a gauge theory identification of stringy degrees of freedom.

%%%%%%%%%%%%%%%%%%%%%%%%
\section*{Acknowledgement}
%%%%%%%%%%%%%%%%%%%%%%%%
We are grateful to J.~R.~David, D.~N.~Kabat and D.~S.~Ponomarev for various illuminating discussions and comments on the manuscript. We also thank B.~Roy for proofreading the manuscript. This work was supported in part by U.S.\ National Science Foundation grants PHY-0855582, PHY11-25915, Department of Energy grant DE-FG02-92-ER40699 and by PSC-CUNY and ERC Self-completion grants.

%%%%%%%%%%%%%%%%%%%%%%%

\end{document}